\documentclass[reprint,amsmath,amssymb,aps,prl]{revtex4-2}
\usepackage[]{xcolor}
\usepackage{graphicx}% Include figure files
\usepackage{dcolumn}% Align table columns on decimal point
\usepackage{bm}% bold math
\usepackage[normalem]{ulem} % for striked-out text

\begin{document}

\preprint{APS/123-QED}

\title{Noise level of a ring laser gyroscope in the femto-rad/s range}

\author{Angela D. V. Di Virgilio$^{1}$}
\author{Francesco Bajardi$^{2,3}$}
\author{Andrea Basti$^{1,4}$}
\author{Nicolò Beverini$^4$}
\author{Giorgio Carelli$^{1,4}$} 
\author{Donatella Ciampini$^{1,4}$}
\author{Giuseppe Di Somma$^{1,4}$}
\author{Francesco Fuso$^{1,4}$}
\author{Enrico Maccioni$^{1,4}$}
\author{Paolo Marsili$^{1,4}$}
\author{Antonello Ortolan$^5$}
\author{Alberto Porzio$^{2,6}$}
    \altaffiliation{Corresponding author alberto.porzio@na.infn.it}
\author{David Vitali$^{7,8}$}

\affiliation{$^1$INFN Sez.~di Pisa, Largo Bruno Pontecorvo 3, I-56127 Pisa, Italy}
\affiliation{$^2$INFN Sez.~di Napoli,
Compl.~Univ.~Monte S.~Angelo, Edificio G, Via Cinthia, I-80126, Napoli, Italy}
\affiliation{$^3$Scuola Superiore Meridionale, Largo San Marcellino 10, I-80138, Napoli, Italy}
\affiliation{$^4$Dipartimento di Fisica, Universit\`a di Pisa,  Largo Bruno Pontecorvo 3, I-56127 Pisa, Italy}
\affiliation{$^5$INFN - LNL , Viale dell’Universit\`a 2, 35020 Legnaro (PD), Italy}
\affiliation{$^6$Department of Civil and Mechanical Engineering -- DICEM, Universit\`{a} di Cassino e Lazio Meridionale, I-03043 Cassino, Italy}
\affiliation{$^7$Physics Division, School of Science and Technology, Universit\`a di Camerino Via Madonna delle Carceri 9, I-62032 Camerino (MC), Italy}
\affiliation{$^8$INFN Sez.~di Perugia, Via A.~Pascoli, I-06123 Perugia, Italy}

\date{\today}% It is always \today, today,

\begin{abstract}
Absolute angular rotation rate measurements with sensitivity better than prad/s would be beneficial for fundamental science investigations. On this regard, large frame Earth based ring laser gyroscopes are top instrumentation as far as bandwidth, long-term operation, and sensitivity are
concerned. Here, we demonstrate that the GINGERINO active-ring laser
%prototype
upper limiting noise is close to $2 \times 10^{-15}$ rad/s for $\sim 2 \times 10^5$ s of integration time, as estimated by the Allan
deviation evaluated in a differential measurement scheme. This is more than a factor 10 better
than the theoretical prediction so far accounted for ideal ring lasers shot–noise with the two beams
counter–propagating inside the cavity considered as two independent propagating modes. This
feature is related to the peculiarity of real ring laser system dynamics that causes phase cross–talking among the two counter–propagating modes. In this context, the independent beam model
is, then, not applicable and the measured noise limit falls below the expected one.
\end{abstract}

\maketitle

Light based interferometers have reached n high level of sensitivity, reliability, and robustness. In most interferometers, two separate beams are injected in two separate paths and recombined to interfere so that differences in path-lengths even smaller than $10^{-14}$ times the wavelength can be resolved \cite{LVK}.

While such measurement scheme is possible thanks to the wave-nature of light, that shows-up as the interference of coherent beams, the corpuscular nature of light sets the intrinsic limit to the sensitivity attainable by interference. This limit is known as shot--noise and it is frequency independent. It intrinsically comes from the stochastic fluctuations in the photon number that, for coherent beams, are Poissonian distributed and so are the detected photo-electrons \cite{Picinbono1970}.
Overcoming the sensitivity limit due to shot-noise is a relevant objective in interferometry and sub-shot noise sensing schemes have been introduced, see e.g. Refs.~\cite{McKenzie2002,Giovannetti2004,Abadie2011,Pradyumna2020}.

Interferometer topologies can be quite different. Paths defined by four mirrors, located at the vertices of a square, define a ring cavity where the two light beams circulates in counter-- and clockwise directions. In this case, the two paths are equal, frequency jitters are negligible, and the interference of the two counter-propagating beams carries information on the non reciprocal effects connected to the direction of circulation. If the frame supporting the four mirrors rotates, the two counter-propagating beams complete the path at different times. In such a configuration, the interference measures the time derivative of the difference in phase acquired by the two beams, rather than the path spatial difference. This feature is the well known Sagnac effect \cite{Sagnac, CRP}.

Sagnac interferometers, in particular their active versions known as Ring Laser Gyroscopes (RLGs), are  commonly used to measure inertial angular rotation (for a review on RLG see Refs. \cite{StedmanRev,Ulli}). Moreover, RLGs with optical cavity area typically above $10\,$m$^2$, when connected to the Earth crust, can be used to measure continuously the absolute angular rotation rate of the Earth, whose value is embedded in the difference in frequency of the two counter-propagating beams. The storage time of the cavity determines the bandwidth of the instrument, typically above $1\,$kHz, and, being the measurement based on frequency reconstruction, the dynamic range is very high: RLGs can detect strong earthquakes and seismological signals in the frequency window $\sim 0.01-30\,$Hz, as well as tiny geodetic signals  in the very low frequency domain ($< 10^{-3}\,$Hz), showing an adequate  sensitivity to probe General Relativity (GR) effects such as the Lense-Thirring and de Sitter \cite{Tartaglia2017}.

Moreover, other non reciprocal effects, related to propagation of the two light beams and connected to the space time structure or symmetries, can be investigated by RLGs. This leads to results relevant in fundamental physics \cite{Scully1981,Capozziello2021,CDR2022} when sensitivity of $5\times 10^{-14}\,$rad/s, or better, are reached, corresponding to 1 part in $10^9$ of the Earth rotation rate.
Sagnac interferometers are also good candidates for investigating the interplay between GR and quantum systems \cite{Fink17,Restuccia19,Toros20,Toros22,Cromb23}.

Since the first model by Cresser et al.~\cite{Cresser82}, developed after concepts described in \cite{Dorschner80}, evaluation of the expected sensitivity limit in actual RLGs \cite{chow,Ulli} has assumed independent counter-propagating beams. This assumption implies neglecting any mechanism that lets the two beams exchange phase information and considering the two laser emission affected by two completely independent phase (Wiener) diffusion processes. So doing, the linewidth of each counter-propagating beam is a Lorentzian and the width of the Sagnac frequency, with a Schawlow–Townes form, is given by the quadratic sum of the two Lorentzians \cite{StedmanRev}.
For GINGERINO \cite{GINGERINO}, a running prototype of the GINGER RLG array \cite{CDR2022,AVS23} located inside the Gran Sasso National Laboratory of INFN, Italy, the model evaluates a shot--noise of about 18 prad/s Hz$^{-1/2}$, taking into account its operating parameters. 

However, in  RLGs the two beams are generated inside the same active medium volume that emits toward the two opposite directions and the laser equations for the two counter-propagating beam amplitudes are coupled to each other \cite{Wilkinson1987}.
In a recent paper, Mecozzi \cite{Mecozzi23} has shown that two of these coupling mechanisms can play a role: back--scattering and gain--scattering, which causes phase locking of the two beams, reducing the effect of phase diffusion in the two modes dynamics. If the Langevin equations for the two beams include them, the noise behaviour of RLGs appears different. Thus, the Allan deviation for the Sagnac signal does not scale as $\sqrt{t}$ (where $t$ is the measurement time), as expected for the shot--noise of interfering independent beams, and the limiting noise falls below the expected shot--noise.

Analysing the data of GINGERINO we had already realised inconsistency with the shot--noise evaluated by the commonly used prediction \cite{PRR2020,EPJC2021,LASER}. This was an indirect evidence, since the analysis focused at low frequency, below 30 Hz, where physical and geophysical investigations are relevant, and  sources of different nature such as, human activity, microseismicity of the crust generated by the ocean, tides and polar motion, temperature and pressure variations, are present.
At low frequency, it is hard to directly measure the shot--noise of an interferometer, and a white noise level can be induced by shocks or electronics. The measurement is, indeed, feasible by subtracting two independent measurements of the same interferometer. In this way the physical signals are subtracted, while the non reciprocal disturbances and the noise of stochastic origin, summed in quadrature, are left. When it is possible to have two independent measurements, the subtraction provides an upper limit for the limiting noise of the interferometer.

In this Letter we report such a differential measurement so giving a conclusive experimental proof that the noise limit of the instrument is well below predictions based on the independent beam model.
The noise floor is found directly by subtracting data obtained from two equivalent beating optical signals at the two outputs of a single beam splitter with no data manipulation, avoiding linear regressions to cancel known signals \cite{Supplemental}.
So doing, we trace-out, in analogy with common mode rejection, all possible rotational signals providing an upper limit for the resulting background noise, including quantum noise sources.
Before presenting and discussing this result, we give a brief overview on the RLG signal and on the measurement methodology and set-up. Later, we give hints on the dominant type of noise for our RLG and discuss the result in terms of the Allan deviation of the time series. 

RLG senses the projection of the angular velocity vector $\vec{\Omega}$ \cite{Symbols} on the area of the closed polygonal cavity. The relationship between the Sagnac pulse frequency $\omega_s$ and the angular rotation rate $\Omega$ reads
\begin{equation}
\omega_s =8 \pi\frac{A}{\lambda L} \Omega \cos{\theta} \;, 
\label{Eq:omegas}
\end{equation}
where $A$ is the area of the cavity, $L$ the perimeter, $\lambda$ the wavelength of the light, and $\theta$ the angle between the area versor $\vec n$  and  $\vec{\Omega}$. Let us now explain in more detail how the signal is extracted from the cavity.

\begin{figure}
    \centering
    \includegraphics[width=8.0cm]{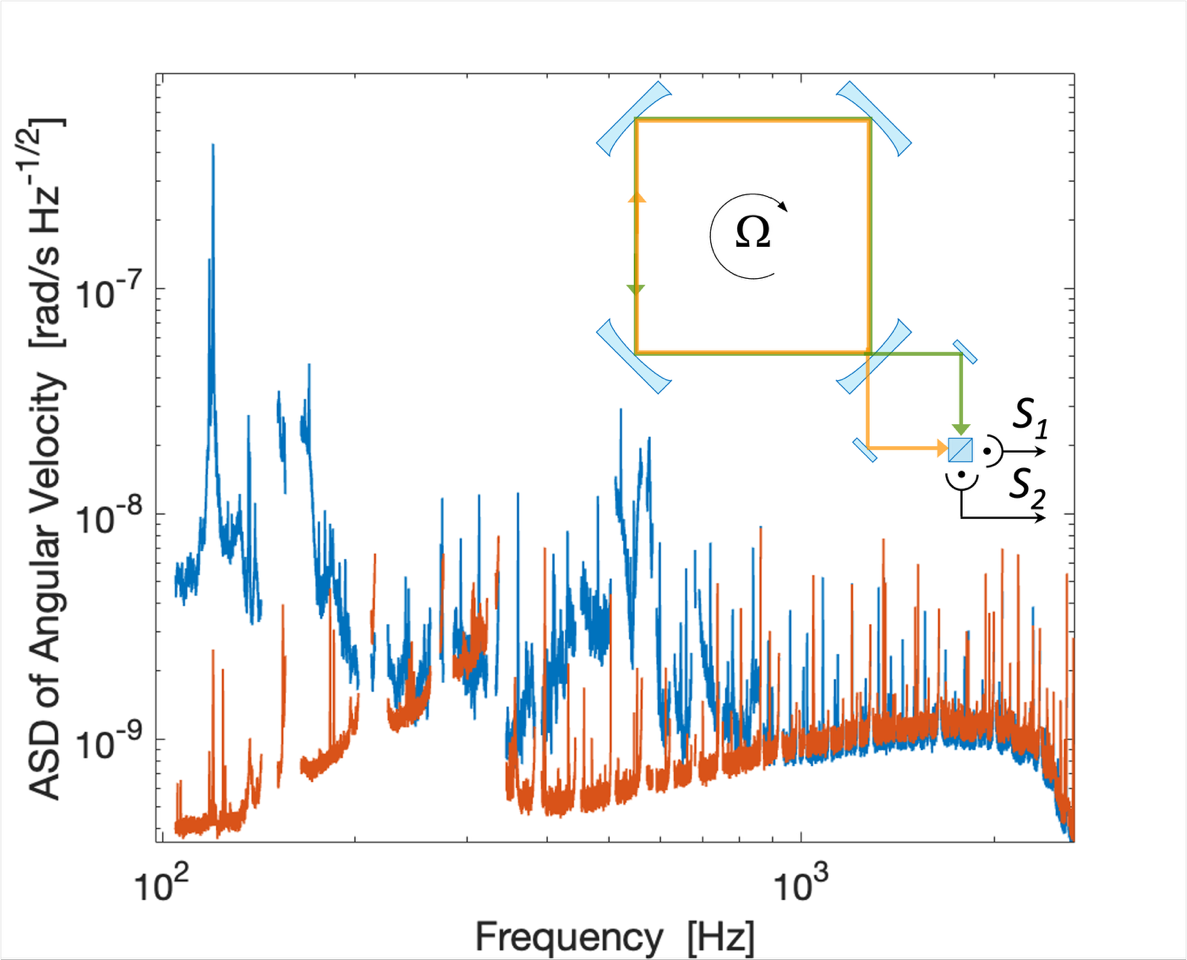}
    \caption{ASD of the angular velocity obtained by considering the difference between either the photodiode signals, $\Omega_d$, in blue, or the corresponding reconstructed frequencies, $\Omega_{n12}/2$, in red. The high frequency range, above 100 Hz, is plotted. Some peaks due to electronics or environmental origin have been removed. The cut--off occurring around 2 kHz is essentially due to the sampling rate. The inset shows a simplified sketch of the RLG setup, including the photodiodes used to produce the $S_1$ and $S_2$ signals.}
    \label{fig:ASD12}
\end{figure}

The two counter-propagating beams transmitted by one of the mirrors at the cavity corners are combined at a beam--splitter, as sketched in the inset of Fig.~\ref{fig:ASD12}. The two resulting mixed beams, observed by two identical photodiodes, embed the measured beat note $\omega_m$.
It contains disturbances induced by the non linear laser dynamics, back--scatter and null--shift,
to be eliminated in order to reach precision and accuracy better than 1 part in $10^5$ \cite{AO2018}. To recover $\omega_s$ we follow a two steps procedure: in the first one, an analytical approach returns $\omega_{s0}$, while the second, more refined one gives $\omega_s$, based on statistical means and on the assumption that the losses of the cavity ($\mu$) change with time.
Corrections between the first and second analysis steps are below the nrad/s range (see \cite{Supplemental} for details on data analysis). 
It has been checked that the result remains valid using raw or refined data; in the present work the first step of the analysis is used ($\omega_{s0}$).
Without loss of generality, it is possible to state that the two photodiode signals $S_i$, with $i = 1,2$, can be expressed as 
\begin{equation}
    S_i = A_g\cdot(-1)^i\cdot\cos\left[{(\omega_s + \omega_n)\cdot t + \phi_n}\right] + V_{n_i}, 
    \label{eq:Si}
\end{equation}
where $A_g$ is a gain factor,
$\omega_n$ indicates the stochastic noise affecting the frequency itself, $\phi_n$ is the stochastic term of the phase, and $V_{n_i}$ is any noise added outside the cavity \cite{diodi}. Here we note that the two signals exiting the two beam-splitter ports are opposite in phase.
The reconstructed frequency signal from each photodiode is  $\omega_i = \omega_s +\omega_{Tn_i}$, where $\omega_{Tn_i}$ takes into account  all noise terms at once, since it is not possible to discriminate among different noise sources, and $\omega_{Tn_i} > \omega_{n}$.
The two interferograms can be used independently or together.
Considering $\omega_i$ associated with $S_i$, and  $\omega_d$ with $S = S_1-S_2$, in $S$ the Sagnac signal is doubled  while the stochastic noise is summed in quadrature, hence $\omega_{d}$ has a signal to noise ratio $\sqrt{2}$ larger than $\omega_i$.
Let us consider $\omega_{n12}=\omega_1-\omega_2$. Calling $\omega_{Tn_d}$ the noise term associated with $\omega_d$, it is straightforward to conclude that $\omega_{n12} \sim 2 \omega_{Tn_d}$. So said, $\omega_{n12}/2$ provides a direct upper noise limit of the apparatus.

Fig.~\ref{fig:ASD12} shows the Amplitude Spectral Distribution (ASD) above $100$ Hz of the angular velocities $\Omega_{d}$ and $\Omega_{n12}/2$, corresponding to $\omega_d$ and $\omega_{n12}$ respectively.
As the rotational
signals are almost absent in the plotted frequency range, the
noise floors are in good agreement with each other, as expected.

\begin{figure}
    \centering
    \includegraphics[width=8.0cm]{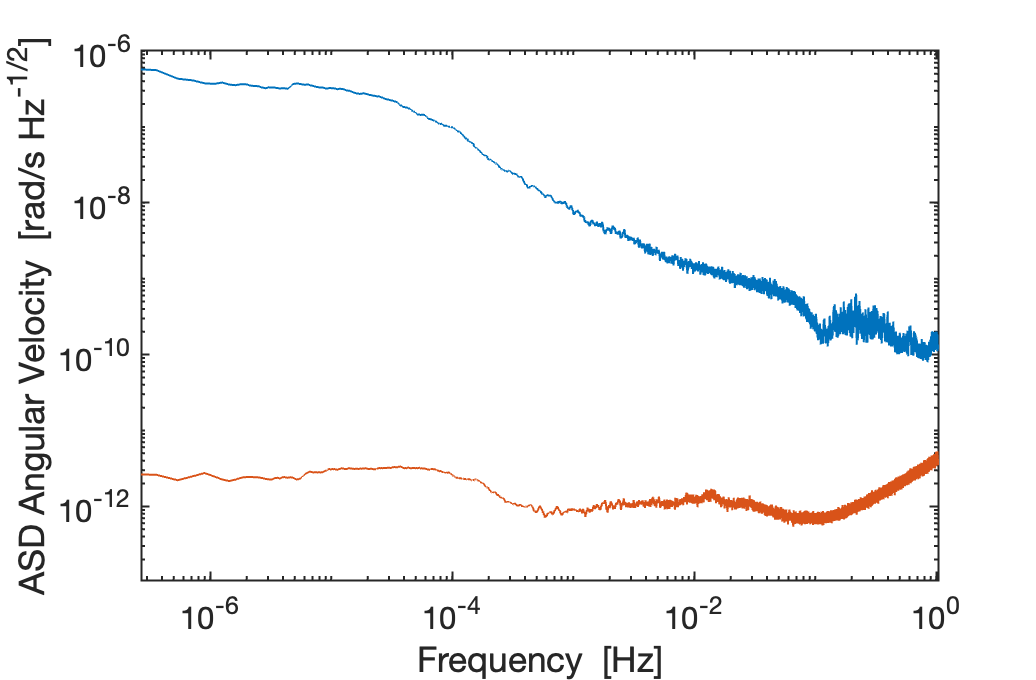}
    \caption{Same of Fig.~\ref{fig:ASD12} for the low frequency range, below 1 Hz, where $\Omega_d$ and $\Omega_{n12}/2$ strongly differ with each other, as expected. In this spectrum two hours of data around a big Mw 5.9 event have been removed (see \cite{Supplemental}); when included, the low frequency bump increases. Contrary to expectations for the shot--noise, $\Omega_{n12}/2$ is not a flat noise and, for GINGERINO, it shows the limit of $2-3\,$prad/s in 1 s measurement time.}
\label{fig:ASDC}
\end{figure}

\begin{figure}
    \centering
    \includegraphics[width=8.0cm]{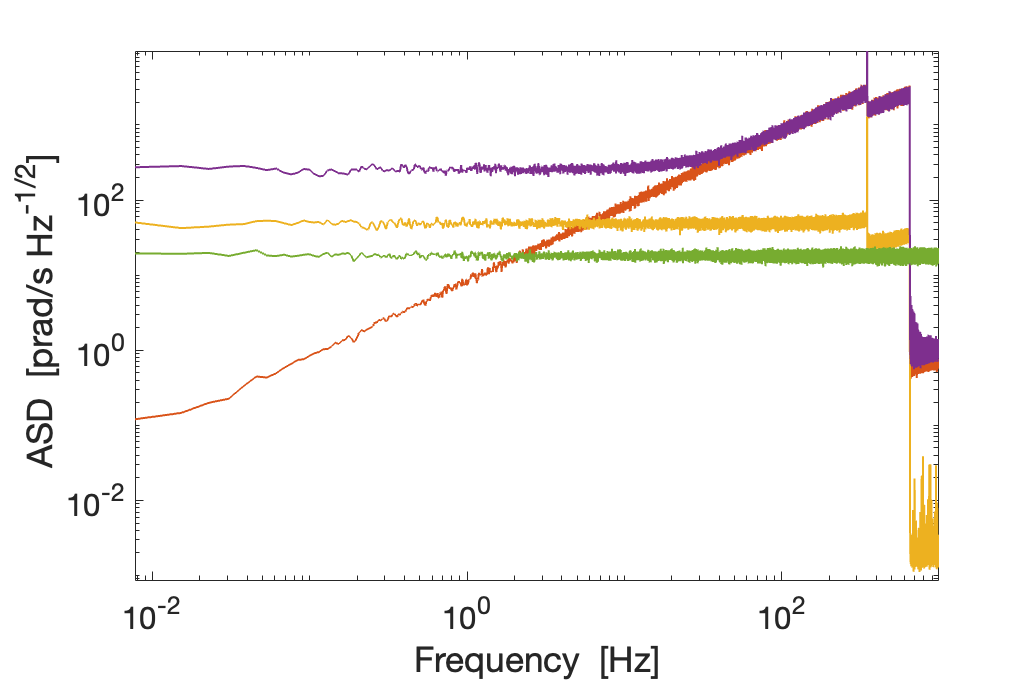}
    \caption{ASDs of the injected frequency noise $\omega_{n}$ (green) and the corresponding reconstructed signal (purple). In red, the ASD of the reconstructed signal obtained by injecting a phase noise $\phi_n$, with $\bar{t} = 0.02$ s integration time, and, in yellow, a Wiener noise $\phi_W$. The injected noise levels correspond to $20\,$prad/s Hz$^{-1/2}$ at $1\,$Hz. We note that the reconstruction procedure at low frequency, for a white frequency noise, returns a noise 20 times higher than the injected one. The discontinuities at the Sagnac Frequency are a feature of the frequency reconstruction and are present also in real data power spectra.}
    \label{fig:my_ASD}
\end{figure}

Figure \ref{fig:ASDC}, instead, compares $\Omega_{d}$ and $\Omega_{n12}/2$ in the low frequency region and it provides the most relevant result of the paper.
Above 0.1 Hz, the latter exhibits  a linear growth whose nature is typical of a phase noise behaviour, as we will also see in
the following (see Fig.~\ref{fig:my_ASD}), and it is flat at lower frequency, with a level around 2 prad/s Hz$^{-1/2}$, a factor 10 below the expected shot--noise that is 18 prad/s Hz$^{-1/2}$.

In order to understand which is the nature of the dominant noise source we have used simulated noise data. Taking into account the general definition of Eq.~\ref{eq:Si}, data have been generated at fixed frequency $\omega_s$ and injecting different stochastic noise sources as $\phi_n$ or $V_n$. These synthetic data have been processed with the same procedure depicted above for real ones.
 
Figure \ref{fig:my_ASD} shows the response of the reconstruction procedure to the injection of three types of noise: white frequency noise ($\omega_n$), white phase noise ($\phi_n$), and phase diffusion noise $\phi_W$ modeled as a Wiener process.
In particular, we report the ASD of the injected noise $\omega_{n}$ and of the corresponding reconstructed signal, as well as the ASD of the reconstructed signal injecting $\phi_{n} = \omega_n \cdot\bar{t}$, with $\bar{t} = 0.02$ s integration time, and $\phi_W$, with standard deviation $12.2$ mrad. 

The contribution of the white stochastic frequency noise  $\omega_n$ is reconstructed by the analysis process as a white noise a factor 20 higher in the low frequency range, then, it grows linearly at higher frequencies. At frequencies above 20 Hz, its behaviour becomes indistinguishable from the one obtained injecting a white phase noise $\phi_n$. The latter produces a power spectrum proportional to frequency over the full frequency span. On the other hand, the phase diffusion noise, simulated as a Wiener process, produces a constant ASD, a factor of 2 higher than the level of the injected noise. It's worth noticing that all ASD of the reconstructed signals show a discontinuity at the Sagnac frequency.

Comparison of the simulated behaviours with that of $\Omega_{n12}/2$ plotted in Fig.~\ref{fig:ASDC} shows that GINGERINO noise, at least above $0.1\;$Hz, is dominated by a phase noise source.

\begin{figure}[t!]
    \centering
    \includegraphics[width=8.0cm]{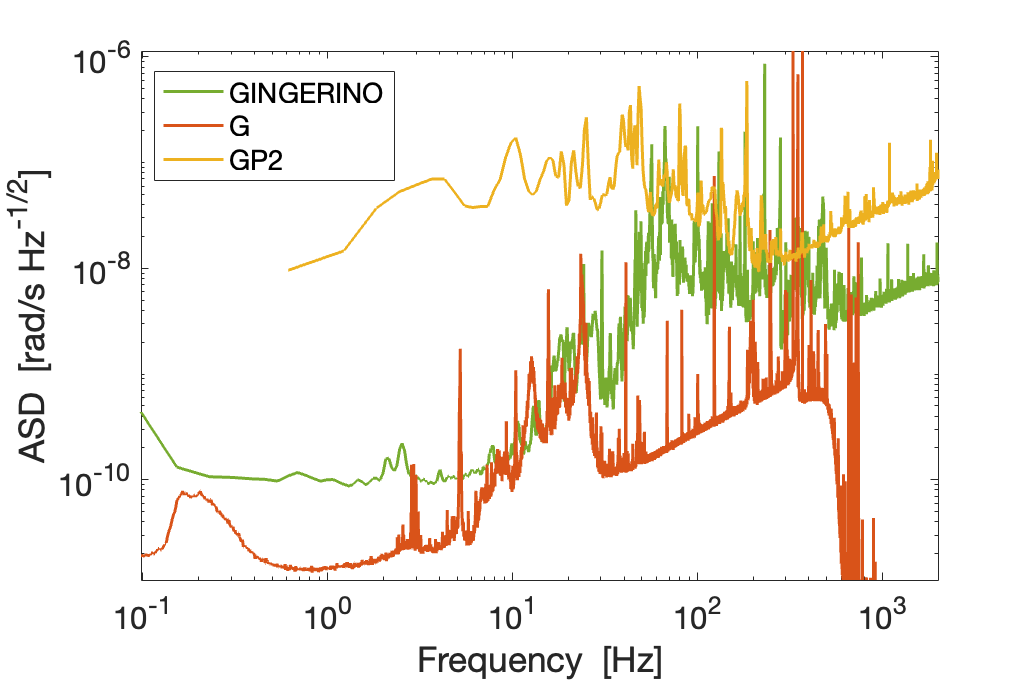}
    \caption{ASDs of angular rotation rate obtained from
$\omega_{s0}$ (single signal evaluation) in rad/s Hz$^{-1/2}$, for G Wettzell, GINGERINO, and GP2. To compare signals from different RLGs we use angular rotation rates instead of rotation velocities (as in Figs. \ref{fig:ASD12} and \ref{fig:ASDC}) that do not take into account the different geometries. The high frequency part of the spectrum shows the phase noise characteristic tail constantly rising with frequency. G, owing to its monolithic structure, is very quiet, GP2 is 1.6 m in side,  and it is located in  a rather noisy environment, that explains the occurrence of a larger noise, and a shorter data set has been used. Data from G are acquired at 2 kHz, its earlier cut-off, occurring around 0.5 kHz is due to the analysis procedure}.
    \label{fig:HNoise}
\end{figure}

To understand if the above is a common feature of different large frame RLGs, we have analysed experimental data produced by four distinct RLGs where a single signal is acquired. As said, the frequency range below 0.1 Hz is affected by laser systematics and contains signals of geophysical origin, thus for this purpose we analyse higher frequencies.

We report in Fig.~\ref{fig:HNoise}  the ASD relative to $\omega_{s0}$ (see \cite{Supplemental} for details) for G--Wettzell \cite{Ulli}, GINGERINO, and GP2 \cite{GP2}, while the ASD of ROMY \cite{ROMY}, that shows a very similar behaviour, is not reported for the sake of clarity.
The minimum of the ASD is in the frequency window $0.1-1\,$Hz, where microseismicity originated by the oceans is present. The region above $5\,$Hz contains regular signals but also a characteristic tail linearly growing with frequency for all RLGs. 
Despite evident differences, due to the different structure and location (GP2 is located in a noisy environment), all three ASDs linearly grow with frequency, already for frequency above few Hz for G and GINGERINO. Comparing this feature with the behaviour obtained with simulated data (see Fig. \ref{fig:my_ASD}) we can conclude that, in these RLGs, phase noise prevails, at least above $0.1\,$Hz, while below the noise level is rather flat, but very low, and further work is required to investigate the origin of this flat noise level.

We stress that the plots of Fig.~\ref{fig:HNoise} do not allow an estimation of the intrinsic noise of the RLG. Indeed, $\omega_{s0}$ contains all the possible signal and noise sources so that the ASD minimal values are biased. On the contrary, the differential detection scheme developed in the present work gives a reliable estimation of the noise floor. In such a case the ASD reports only noise sources, quantum noise included, that independently affect the two outputs of the beam splitter.

\begin{figure}
    \centering
   \includegraphics[width=8.0cm]{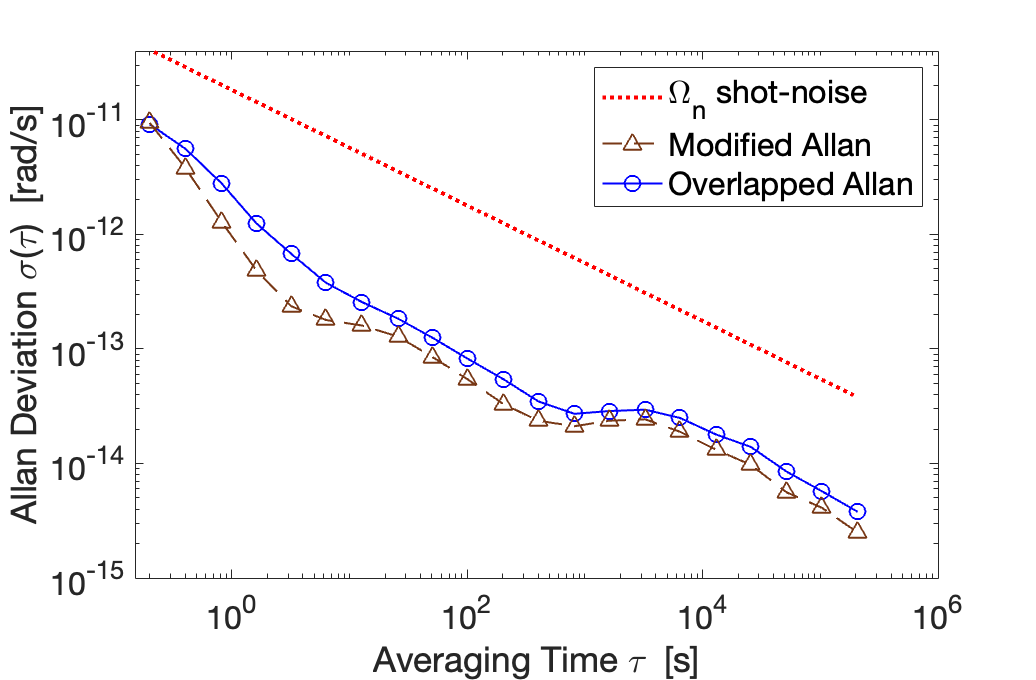}
    \caption{Overlapped and Modified Allan Deviation of $\Omega_{n12}/2$ expressed in rad/s. Plotted data have been obtained by using STABLE32 \cite{STABLE32}. The red dashed line represents the expected shot--noise for GINGERINO using the independent beam model. 
    The points at 1 Hz give values in the range of 2 prad/s in agreement with the ASD presented in Fig.~\ref{fig:ASDC}. Then, the Modified Allan Deviation reaches the value of $2.1\pm 0.01$ frad/s in 2.5 days of integration time, that corresponds to $4.3\times 10^{-11}$ the Earth rotation rate.}
    \label{fig:highNoise}
\end{figure}

In order to systematically compare the sensitivity reached in GINGERINO with the estimation of its noise limit given by the independent beam model, we used Allan deviation of the time series.
We report in Fig.~\ref{fig:highNoise} the Overlapped and Modified Allan Deviations; for sake of completeness, we remark that the Overlapped Allan Deviation,  evaluated by using $\omega_{m2}-\omega_{m1}$, provides very similar results \cite{Supplemental}. The GINGERINO noise drops to $4$ and $2.63$ frad/s in approximately 2.4 days of integration time, respectively, corresponding to $1.23$ and $1.87$ in $10^{10}$ the Earth rotation rate, a level sufficient for detecting fundamental physics effects with an array of RLGs \cite{CDR2022,Capozziello2021}. In the plot, the red-dashed line represents the shot-noise level expected for GINGERINO using the independent beam model.

It is proved that, below $0.1$ Hz, the large RLG
GINGERINO shows a limiting noise floor in the prad/s Hz$^{-1/2}$ range, well below what expected for the shot--noise in this type of apparatus taking for granted the independent beam model \cite{Cresser82}. This experimental noise limit has been obtained by subtracting two independent rotation signals. These signals come from the two outputs of a single beam--splitter placed at one of the cavity corners. So doing, the estimated noise level represents an upper limit to the inherent noise affecting the apparatus.
While this experimental finding suggests that a complete model of the system should take into account the complex interdependent dynamics of the counter-propagating beams, it gives a conclusive proof of the feasibility of fundamental physics measurements once an array of RLGs is available.

In the complex dynamics of RLG, there are physical mechanisms whose effect is to couple the two beam dynamics, so establishing some sort of correlation in time that an independent beam model cannot account for. As suggested by Mecozzi in \cite{Mecozzi23}, locking-mechanisms decouple ``noise of the beat note from the frequency noise of the individual modes, thus allowing the realization of sub-shot--noise laser gyros''.

A full quantitative comprehension of the actual limit to RLG noise goes beyond the scope of the present paper and requires a detailed model that considers both locking mechanisms in a unified frame, that could be applied to a large range of frequencies.

This result paves the way to the use of high sensitivity RLG in general relativity and beyond, as well as quantum physics research where tiny effects are expected \cite{Frontiers24}.

\section*{Acknowledgement}
The authors thank K. U. Schreiber, J. Kodet, H. Igel and A. Brotzer for providing the data of G Wettzell and ROMY to investigate the high frequency noise.


\begin{thebibliography}{99}

\bibitem{LVK}
Abbott, B.P., \textit{et al.}, Living Rev. Relativ.  {\bf 23}, 3 (2020).
\bibitem{Picinbono1970}
B. Picinbono, C. Bendjaballah, and J. Pouget, J. Mat. Phys. {\bf 11}, 2166 (1970).
\bibitem{McKenzie2002}
Kirk McKenzie, Daniel A. Shaddock, David E. McClelland, Ben C. Buchler, and Ping Koy Lam, Phys. Rev. Lett. \textbf{88}, 231102 (2002).
\bibitem{Giovannetti2004}
Vittorio Giovannetti, Seth Loyd, and Lorenzo Maccone, Science, \textbf{306}, 1330--1336 (2004).
\bibitem{Abadie2011}
The LIGO collaboration, Nature Phys. \textbf{7}, 962–-965 (2011).
\bibitem{Pradyumna2020}
S. T. Pradyumna, \textit{et al.}, Commun. Phys. \textbf{3},104 (2020).
\bibitem{Sagnac}
G. Sagnac, Comptes Rendus {\bf 157}, 708 (1913); {\bf 157}, 1410 (1913).
\bibitem{CRP}
The Sagnac effect: 100 years later / L'effet Sagnac: 100 ans après
Special Issue edited by Gauguet Alexandre, Compte Rendus Physique, {\bf 15}, Issue 10, (2014).
\bibitem{StedmanRev}
G.E. Stedman, Rep. Prog. Phys. \textbf{60}, 615 (1997).
\bibitem{Ulli}
K.U. Schreiber and K.U. Wells, Rev. Sci. Instrum. {\bf 84}, 041101 (2013).
\bibitem{Tartaglia2017}
A. Tartaglia, A.D.V. Di Virgilio, J. Belfi, N. Beverini, and M.L. Ruggiero, Eur. Phys. J. Plus {\bf 132}, 73 (2017).
\bibitem{Scully1981}
M.O. Scully, M.S. Zubairy, and M.P. Haugan, Phys. Rev. A {\bf 24}, 2009 (1981).
\bibitem{Capozziello2021}
S. Capozziello, \textit{et al.}, Eur. Phys. J Plus, {\bf 136}, 394 (2021).
\bibitem{CDR2022}
C. Altucci, \textit{et al.}, Math. Mech. Complex Syst. {\bf 11}, 203--234 (2023).
\bibitem{Fink17}
M. Fink, \textit{et al.}, Nat. Commun. {\bf 8}, 1 (2017).
\bibitem{Restuccia19}
S. Restuccia, M. Toro\v{s}, G. M. Gibson, H. Ulbricht, D. Faccio, and M. J. Padgett, Phys. Rev. Lett. {\bf 123},110401 (2019).
\bibitem{Toros20}
M. Toro\v{s}, S. Restuccia, G. M. Gibson, M. Cromb, H. Ulbricht, M. Padgett, and D. Faccio, Phys. Rev. A {\bf 101}, 043837 (2020).
\bibitem{Toros22}
M. Toro\v{s}, M. Cromb, M. Paternostro, and D. Faccio, Phys. Rev. Lett. {\bf 129}, 260401 (2022).
\bibitem{Cromb23}
M. Cromb, S. Restuccia, G. M. Gibson, M. Toro\v{s}, M. J. Padgett, and D. Faccio, Phys. Rev. Res. {\bf 5}, L022005 (2023).
\bibitem{Cresser82}
J.D. Cresser, W.H. Louisell, P. Meystre, W. Schleich, and M.O. Scully, Phys. Rev. A {\bf 25}, 2214 (1982); J.D. Cresser, D. Hammonds, W.H. Louisell, P. Meystre, and H. Risken, Phys. Rev. A {\bf 25}, 2226 (1982); J.D. Cresser, Phys. Rev. A {\bf 26}, 398 (1982).
\bibitem{Dorschner80}
T. Dorschner, H. Haus, M. Holz, I. Smith, and H. Statz, IEEE J. Quantum Electron. {\bf 16}, 1376 (1980).
\bibitem{chow}
W.W. Chow, J. Gea-Banacloche, L.M. Pedrotti, V.E. Sanders, W. Schleich, and M.O. Scully, Rev. Mod. Phys. {\bf 57}, 61 (1985).
\bibitem{GINGERINO}
Jacopo Belfi, \textit{et al.}, Rev. Sci. Instrum. {\bf 88}, 034502 (2017).
\bibitem{AVS23}
C. Altucci, \textit{et al.}, AVS Quantum Sci. \textbf{5}, 045001 (2023).
%\bibitem{MEMOCS23}
%C. Altucci, \textit{et al.}, MEMOCS \textbf{11}:203–234, (2023);
\bibitem{Wilkinson1987}
J.R. Wilkinson, Prog. Quant. Electr. {\bf 11}, 1 (1987).
\bibitem{Mecozzi23}
Antonio Mecozzi, Optica {\bf 10}, 1102-1110 (2023).
\bibitem{PRR2020} A.D.V. Di Virgilio, \textit{et al.}, Phys. Rev. Res. {\bf 2}, 032069(R) (2020).
\bibitem{EPJC2021}
A.D.V. Di Virgilio, \textit{et al.}, Eur. Phys. J. C {\bf 81}, 400 (2021); A. Basti, \textit{et al.}, Eur. Phys. J. Plus, \textbf{136}, 537, (2021).
\bibitem{LASER}
A.D.V. Di Virgilio, N. Beverini, G. Carelli, D. Ciampini, F. Fuso, and E. Maccioni, Eur. Phys. J. C {\bf 79}, 573 (2019); A.D.V. Di Virgilio, N. Beverini, G. Carelli, D. Ciampini, F. Fuso, U. Giacomelli, E. Maccioni, and A. Ortolan, Eur. Phys. J. C {\bf 80}, 163 (2020).
\bibitem{Supplemental}
See Supplemental Materials for details. 
\bibitem{Symbols}
A table of all the relevant symbols used in the paper is included in the Supplemental Materials. In particular, capital $\Omega$, given in rad/s, will always indicate angular rotation rates while small $\omega$, measured in Hz, will indicate electrical signal pulse frequencies experimentally obtained from the time dependent Sagnac interferograms.
\bibitem{AO2018}
J. Belfi, N. Beverini, G. Carelli, A. Di Virgilio, U. Giacomelli, E. Maccioni, A. Simonelli, F. Stefani, and G. Terreni, Appl. Opt.  {\bf 57}, 5844--5851 (2018).
\bibitem{diodi}
The last term includes noise equivalent power of the photodiode, or amplitude fluctuations of the light at the detector, etc.
\bibitem{GP2}
E. Maccioni, N. Beverini, G. Carelli, G. Di Somma, A. Di Virgilio, and P. Marsili, Appl. Opt. {\bf 61}, 9256--9261 (2022).
\bibitem{ROMY}
H. Igel, \textit{et al.}, Geophys. J. Int. {\bf 225}, 684--698 (2021).
\bibitem{STABLE32}
STABLE 32 is a program for frequency stability analysis freely available at: http://www.stable32.com/.
\bibitem{Frontiers24}
Francesco Giovinetti, \textit{et al.}, Front. Quantum Sci. Technol. \textbf{3} (2024); doi:10.3389/frqst.2024.1363409.

\end{thebibliography}
\end{document}